%% file: 00-main.tex
\documentclass[sigconf,review=false,anonymous=false]{acmart}
\usepackage{booktabs}
\usepackage{amsmath}
\usepackage{amsfonts}
\usepackage{amssymb}
\usepackage[inline]{enumitem}
\usepackage{subfigure}
\usepackage{array}
\usepackage{censor}

\copyrightyear{2018}
\acmYear{2018}
\setcopyright{acmcopyright}
\acmConference[SIGIR '18]{The 41st International ACM SIGIR Conference on Research \& Development in Information Retrieval}{July 8--12, 2018}{Ann Arbor, MI, USA}
\acmBooktitle{SIGIR '18: The 41st International ACM SIGIR Conference on Research \& Development in Information Retrieval, July 8--12, 2018, Ann Arbor, MI, USA}
\acmPrice{15.00}
\acmDOI{10.1145/3209978.3210127}
\acmISBN{978-1-4503-5657-2/18/07}

\newcommand{\eg}{\emph{e.g.}}
\newcommand{\wrt}{\emph{w.r.t. }}

\makeatletter
\g@addto@macro\normalsize{%
  \setlength\abovedisplayskip{2pt}
  \setlength\belowdisplayskip{3pt}
  \setlength\abovedisplayshortskip{2pt}
  \setlength\belowdisplayshortskip{3pt}
}

\newif\ifanon
\anonfalse

\begin{document}
\title{Optimizing Query Evaluations Using Reinforcement Learning for Web Search}

\author{Corby Rosset, Damien Jose, Gargi Ghosh, Bhaskar Mitra, and Saurabh Tiwary}
\affiliation{%
  \institution{Microsoft AI \& Research}
}
\email{{corosset, dajose, gghosh, bmitra, satiwary}@microsoft.com}

\begin{abstract}
In web search, typically a candidate generation step selects a small set of documents---from collections containing as many as billions of web pages---that are subsequently ranked and pruned before being presented to the user. In
\ifanon
\else
Bing,
\fi
the candidate generation involves scanning the index using statically designed match plans that prescribe sequences of different match criteria and stopping conditions. In this work, we pose match planning as a reinforcement learning task and observe up to 20\% reduction in index blocks accessed, with small or no degradation in the quality of the candidate sets.
\end{abstract}



\maketitle
\input{01-intro}
\input{02-related}
\input{03-prelim}
\input{04-model}
\input{05-method}
\input{06-results}
\input{07-conclusion}


\bibliographystyle{ACM-Reference-Format}
\bibliography{bibtex.bib} 

\end{document}

%% file: 01-intro.tex
\section{Introduction}
\label{sec:intro}

In response to short text queries, search engines attempt to retrieve the top few relevant results by searching through collections containing billions of documents \citep{van2016estimating}, often under a second \citep{teevan2013slow}. To achieve such short response times, these systems typically distribute the collection over multiple machines that can be searched in parallel \citep{croft2010search}. Specialized data structures---such as inverted indexes \citep{witten1999managing, zobel2006inverted}---are used to identify an initial set of candidates that are progressively pruned and ranked by a cascade of retrieval models of increasing complexity \citep{matveeva2006high, wang2011cascade}. The index organization and query evaluation strategies, in particular, trade-off retrieval effectiveness and efficiency during the candidate generation stage. However, unlike in late stage re-ranking where machine learning (ML) models are commonplace \citep{qin2010letor, Liu:2009}, the candidate generation frequently employs traditional retrieval models with few learnable parameters.

In 
\ifanon
\else
Bing,
\fi
the document representation consists of descriptions from multiple sources---popularly referred to as \emph{fields} \citep{robertson2004simple, zamani2017neural}. 
\ifanon
\else
Bing
\fi
maintains an inverted index per field, and the posting list corresponding to each term may be further ordered based on document-level measures \citep{long2003optimized}, such as
\ifanon
\else
\emph{static rank} \citep{richardson2006beyond}.
\fi
During query evaluation, the query is classified into one of few pre-defined categories, and consequently a \emph{match plan} is selected. Documents are scanned based on the chosen match plan which consists of a sequence of \emph{match rules}, and corresponding stopping criteria. A match rule defines the condition that a document should satisfy to be selected as a candidate for ranking, and the stopping criteria decides when the index scan using a particular match rule should terminate---and if the matching process should continue with the next match rule, or conclude, or reset to the beginning of the index. These match plans influence the trade-off between how quickly
\ifanon
\else
Bing
\fi
responds to a query, and its result quality.  E.g., long queries with rare intents may require more expensive match plans that consider the body text of the documents, and search deeper into the index to find more candidates. In contrast, for a popular navigational query a shallow scan against a subset of the document fields---\eg, URL and title---may be sufficient. Prior to this work, these match plans were hand-crafted and statically assigned to each query category in
\ifanon
\else
Bing.
\fi

We cast match planning as a reinforcement learning (RL) task. We learn a policy that sequentially decides which match rules to employ during candidate generation. The model is trained to maximize a cumulative reward computed based on the estimated relevance of the additional documents discovered, discounted by their cost of retrieval. We use table-based Q-learning and observe significant reduction in the number of index blocks accessed---with small or no degradations in the candidate set quality.

%% file: 02-related.tex
\section{Related work}
\label{sec:related}

Response time is a key consideration in web search. Even a 100ms latency has been shown to invoke negative user reactions \citep{brutlag2009speed, schurman2009performance}. A large body of work in information retrieval (IR) has, therefore, focused on efficient query evaluations---\eg, \citep{broder2003efficient, dean2009challenges, hawking2017efficiency}. In the context of machine learning based approaches to retrieval, models have been proposed that incorporate efficiency considerations in feature selection \citep{wang2010ranking, wang2010learning}, early termination \citep{cambazoglu2010early}, and joint optimization \citep{wang2011cascade}. Predicting query response times has been explored for intelligent scheduling \citep{macdonald2012learning}, as well as models for aggressive pruning \citep{tonellotto2013efficient, culpepper2016dynamic, yun2015optimal}. Finally, reinforcement learning has been applied in general to information retrieval \citep{nogueira2017task} and extraction \citep{narasimhan2016improving} tasks. However, we believe this is the first work that employs reinforcement learning for jointly optimizing efficiency and performance of query evaluation.

%% file: 03-prelim.tex
\section{Preliminaries}
\label{sec:prelim}

\begin{figure*}
\includegraphics[width=.95\textwidth]{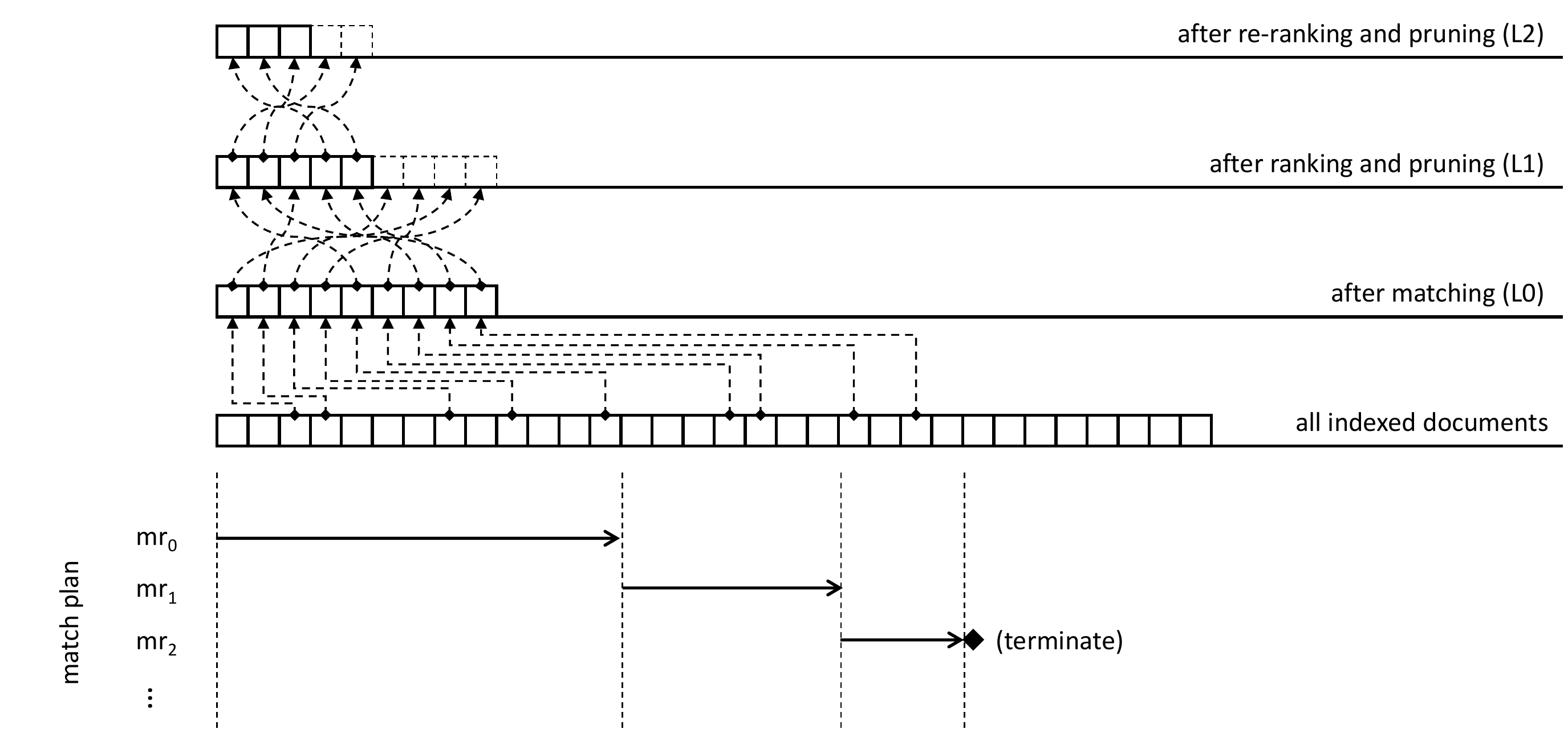}
\caption{A telescoping architecture employed in
\ifanon
\else
Bing's
\fi
retrieval system. Documents are scanned using a pre-defined match plan. Matched documents are passed through additional rank-and-prune stages.}
\label{fig:stack}
\end{figure*}

\ifanon
\else
\paragraph{Web scale retrieval at Bing}
\fi
We focus on the problem of efficient candidate generation for web search. We perform our experiments on top of the production system deployed at
\ifanon
\else
Bing.
\fi
We briefly describe this baseline system in this section. To avoid disclosing any proprietary details about the design of
\ifanon
\else
Bing
\fi
search we only include the information relevant to our evaluation setup.

\ifanon
\else
Bing
\fi
employs a telescoping framework \citep{matveeva2006high} to iteratively prune the set of candidate documents considered for a query. On receiving a search request, the backend classifies the query based on a set of available features---the historical popularity of the query, the number of query terms, and the document frequency of the query terms---into one of the few pre-determined categories. Based on the query category, a match plan---comprising of a sequence of match rules $\{mr_0 \ldots mr_l\}$---is selected that determines how the index should be scanned. Each match rule specifies a criteria that is used to decide whether a document should be included as a candidate. A query may have multiple terms and a document may be represented in the index by multiple fields. A typical match rule comprises of a conjunction of the query terms that should be matched, and for each query term a disjunction of the document fields that should be reviewed. For example, for the query ``halloween costumes'' a match rule $mr_A \rightarrow (\text{halloween} \in A|U|B|T)\wedge (\text{costumes} \in A|U|B|T)$ may specify that each term must have a match in at least one of the four document fields---anchor text (A), URL (U), body (B), or title (T). For the query ``facebook login'', in contrast, a different match rule $mr_B \rightarrow (\text{facebook} \in U|T)$---that only considers the URL and the title fields, and relaxes the matching constraint for the term ``login''---may be more appropriate. While $mr_A$ may uncover more candidates by matching against additional fields, $mr_B$ is likely to be faster because it spends less time analyzing each document. If we assume that the index is sorted by
\ifanon
\else
static rank,
\fi
then $mr_B$ is still likely to locate the right document satisfying the navigational intent.

In
\ifanon
\else
Bing,
\fi
the index data is read from disk to memory in fixed sized contiguous blocks. As the match plan is executed, two accumulators keep track of the number of blocks accessed $u$ from disk and the cumulative number of term matches $v$ in all the inspected documents so far. The match plan uses these counters to define the stopping condition for each of the match rules. When either of the counters meet the specified threshold, the match rule execution terminates. Then, the match plan may specify that the scanning should continue with the next match rule, or the search should terminate. The match plan may also choose to reset the scan to the beginning of the index before continuing with the next match rule.

After the match plan execution terminates, the selected candidates are further ranked and pruned by a cascade of machine learning models. Figure \ref{fig:stack} visualizes this telescoping setup. The matching stage---referred to as level 0, or L0---is followed by a number of rank-and-prune steps (\eg, L1 and L2). This telescoping setup typically runs on each individual machine that has a portion of the document index, and the results are aggregated across all the machines, followed by more rank-and-prune stages. A significant amount of literature exists on machine learning approaches to ranking \citep{Liu:2009, mitra2017introduction}. In this work, we instead study the application of reinforcement learning to the matching stage.

\paragraph{Desiderata of candidate generation}

The candidate generation has a strong influence on both the quality of
\ifanon
\else
Bing's
\fi
results, as well as its response time. If the match plan fails to recall relevant candidates, the ranking stages that follow have no means to compensate for the missing documents. Therefore, the match plan has to draw a balance between the cost and the value of performing more sophisticated query-document analysis (\eg, considering additional document fields). Constructing a match plan that performs reliably on a large number of distinctly different queries classified under the same category is a difficult task. A reasonable alternative may be to learn a policy that adapts the matching strategy at run-time based on the current state of the candidate generation process. Therefore, we learn a policy that sequentially selects matching rules based on the current state---or decides to terminate or reset the scan. Notably, in reinforcement learning this approach is similar to an agent choosing between $k$ available actions based on its present state.

In the telescoping setup, it is important for the matching function to select documents that are likely to be ranked highly by the subsequent models in the pipeline. This means given a choice between two documents with equal number of query term matches, the match plan should surface the document that the rankers in stage L1, and above, prefer. In Section \ref{sec:model}, we will describe our reward function which uses the L1 scores as an approximation of the document's relevance. This implicitly optimizes for a higher agreement between our matching policy and upstream ranking functions.

Finally, it is desirable that our matching strategy is customized for each query category. For example, the optimal matching policy for long queries containing rare terms is unlikely to be the best strategy for short navigational queries. We, therefore, train separate policies for each query category.

%% file: 04-model.tex
\section{Reinforcement Learning for Dynamic Match Planning}
\label{sec:model}

In reinforcement learning, an agent selects an action $a \in \mathcal{A}$ based on the current state $s \in \mathcal{S}$. In response, the environment $E$ provides an immediate reward $r(s, a)$ and a new state $s'$ to the agent. The transition to $s'$ is usually stochastic, and the goal of the agent is to maximize the expected cumulative long-term reward $R$, which is the time-discounted sum of immediate rewards.

\begin{align}
R &= \sum_{t=0}^T \gamma^{t} r(s_t, a_t)\quad,\quad 0 < \gamma \leq 1
\end{align}

where, $\gamma$ is the discount rate. The goal of the agent is to learn a policy $\pi_{\theta} : \mathcal{S} \rightarrow \mathcal{A}$ which maximizes the cumulative discounted reward $R$. In our setup, the action space includes the choice of 
\begin{enumerate*}[label=(\roman*)]
  \item the $k$ different match rules,
  \item resetting the scan to the beginning of the index, or
  \item terminating the candidate generation process.
\end{enumerate*}

\begin{align}
\mathcal{A} = \{\text{mr}_1, \ldots, \text{mr}_k\} \cup \{a_{\text{reset}}, a_{\text{stop}}\}
\end{align}

Our state $s_t \in \mathcal{S}$ is a function of the cumulative index blocks accessed $u_t$ and the cumulative number of term matches $v_t$ at time $t$. We implement table based Q-learning \citep{watkins1992q} which requires that the state space to be discrete. So, we run the baseline match plans from
\ifanon
\else
Bing's
\fi
production system and collect a large set of $\{u_t, v_t\}$ pairs recording after every match rule execution. We assign these points to $p$ bins, such that each bin has roughly the same number of points. These $p$ bins serves as our discrete state space.

During training, we want to reward a policy $\pi_{\theta}$ for choosing an action $a$ at state $s_t$ that maximizes the total estimated relevance of the documents recalled, while minimizing the index blocks accessed. So, our reward function has the following form:

\begin{align}
r_\text{agent}(s_t,a_t) = \frac{\sum_{i}^{m_{t+1}}{g(d_i)}}{n \cdot u_{t+1}}
\end{align}

$g(d_i)$ is the relevance of the $i^\text{th}$ document which we estimate based on the L1 ranker score from the subsequent level of our telescoping setup. The constant $n$ determines the number of top ranked documents we consider in the reward computation, where the ranking is determined by the L1 model. The $u_{t+1}$ component in the denominator penalizes the model for additional documents inspected. The final reward is computed as the difference between the agent's reward and the reward achieved by executing the production baseline match plan:

\begin{equation}
r(s, a) = r_{\text{agent}}(s, a) - r_{\text{production}}(s, a)
\end{equation}

If no new documents are selected, we assign a small negative reward. At test time, we greedily select the action with the highest predicted Q-value. The index scan is terminated when the policy chooses $a_\text{stop}$, or we surpass a maximum execution time threshold.

%% file: 05-method.tex
\section{Data and Experiments}
\label{sec:method}

To train our model, we sample approximately one million queries from
\ifanon
\else
Bing's
\fi
query logs. We train our policies individually for each query category using the corresponding queries from this sampled dataset. We set the size of our state space $p$ to 10K, and during training inspect the top five ($n=5$) documents for computing the reward. For evaluation, we use two query sets---one generated by uniformly sampling from the set of distinct queries in
\ifanon
\else
Bing's
\fi
query log (unweighted set), and the other using a sampling probability that is proportional to the historical popularity of the query (weighted set). For each query, we have a number of documents that have been previously rated using crowd-sourced annotators on a five-point relevance scale.

\ifanon
\else
Bing's
\fi
index is distributed over a large number of machines. We train our policy using a single machine---containing one shard of the index---but test against a small cluster of machines containing approximately $10\%$ of the entire index. During evaluation, the same policy is applied on every machine which, however, may lead to executing different sequences of match rules on each of them.

\paragraph{Metrics} We compare the candidate sets generated by the baseline match plans and our learned policies \wrt both relevance and efficiency. Each candidate set $D$ is unordered because it precedes the ranking steps. To quantify the relevance of an unordered candidate set using graded relevance judgments, we use the popular NDCG metric but without any position based discounting. We compute the Normalized Cumulative Gain (NCG) for $D$ as follows:

\begin{align}
\text{CumGain} &= \sum _{i=1}^{|D|}{\text{gain}_i} \\
\text{NCG} &= \frac{\text{CumGain}}{\text{CumGain}_\text{ideal}}
\end{align}

We limit $|D|$ to 100, and average the NCG values over all the queries in the test set. To measure efficiency, we consider the number of index blocks accessed $u$ during the index scan. In our experiments, any reduction in $u$ show a linear relationship with reduction in the execution time of the candidate generation step. Unfortunately, we can not report these improvements in execution time due to the confidential nature of such measurements.

%% file: 06-results.tex
\section{Results}
\label{sec:results}

\begin{table}
\begin{center}
\caption{Changes in NCG and the index blocks accessed $u$ from our learned policy relative to production baselines. In both categories, we observe significant reduction in index blocks accessed, although at the cost of some loss in relevance in case of CAT1. All the differences in NCG and $u$ are statistically significant ($p < 0.01$). Coverage of CAT2 queries in the unweighted set is too low to report numbers.}
\label{tbl:results}
\begin{tabular}{l>{\centering\arraybackslash}m{1.7cm}>{\centering\arraybackslash}m{1.2cm}>{\centering\arraybackslash}m{1.9cm}}
  \toprule
 & Segment size & NCG$\mathcal{@}100$ & Index block accessed \\
 \midrule
\multicolumn{4}{l}{\textbf{CAT1}}\\ 
Weighted set & 7.2\% & -1.8\% & -17.5\% \\
Unweighted set & 3.2\% & -6.2\% & -16.3\% \\
 \midrule
\multicolumn{4}{l}{\textbf{CAT2}}\\ 
Weighted set & 10.1\% & +0.2\% & -22.7\% \\
Unweighted set & <1\% & - & - \\
  \bottomrule
\end{tabular}
\end{center}
\end{table}

At the time of writing this paper, we have experimented with two of the query categories. CAT1 consists of short multi-term queries with few occurrences over last 6 months. CAT2 includes multi-term queries, where every term has moderately high document frequency. As the absolute numbers are confidential, we report the relative improvements against the
\ifanon
\else
Bing
\fi
production system in Table~\ref{tbl:results}. Notably, these efficiency improvements---also highlighted in Figure~\ref{fig:plot}---are over a strong baseline that has been tuned continuously by many 
\ifanon
\else
Bing
\fi
engineers over several years.

%% file: 07-conclusion.tex
\section{Discussion and Conclusions}
\label{sec:conclusion}

Many recent progresses in IR have been fueled by new machine learning techniques. ML models are typically slower and consume more resources than traditional IR models, but can achieve better retrieval effectiveness by learning from large datasets. Better relevance in exchange for few additional milliseconds of latency may sometimes be a fair trade. But we argue that machine learning can also be useful for improving the speed of retrieval. Not only do these translate into material cost savings in query serving infrastructure, but milliseconds of saved run-time can be re-purposed by upstream ranking systems to provide better end-user experience.

\begin{figure}[t]
\includegraphics[width=.95\linewidth]{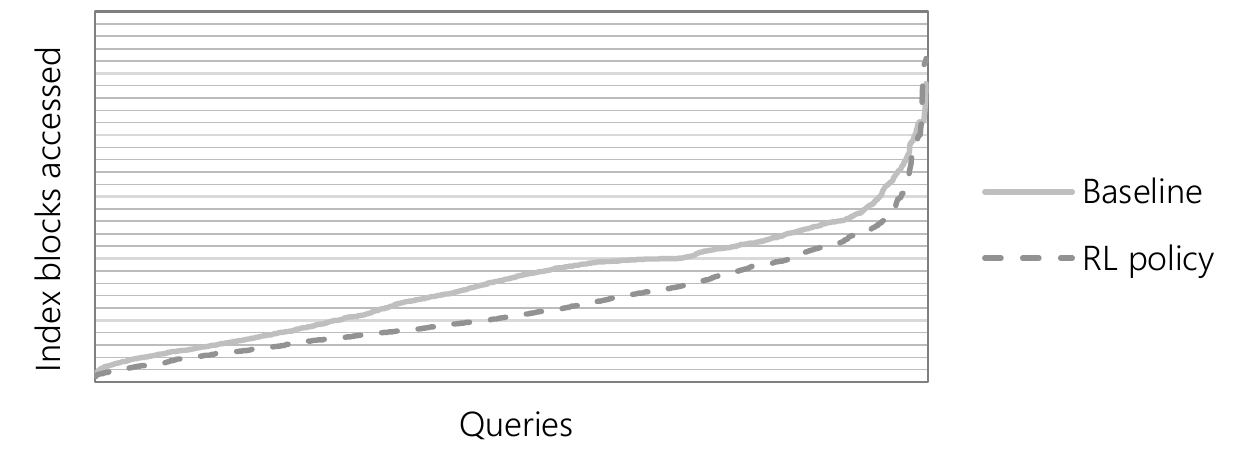}
\caption{The reduction in index blocks accessed from the learned policy for CAT2 queries on the weighted set. We intentionally leave out the actual page access numbers on the $y$-axis because of confidentiality. The queries on the $x$-axis are sorted by page access independently for each treatment.}
\label{fig:plot}
\end{figure}